\documentclass[acmsmall]{acmart}

\usepackage{wrapfig}
\usepackage{fancyhdr}
\usepackage{subcaption}
\usepackage{graphicx} 

\AtBeginDocument{%
  }

\setcopyright{none}
\copyrightyear{2018}
\acmYear{2018}
\acmDOI{XXXXXXX.XXXXXXX}

\acmSubmissionID{188}
\renewcommand\footnotetextcopyrightpermission[1]{}
\begin{document}
\title{Cultivating Multidisciplinary AI Workforce Development on iTiger GPU Cluster: Practices and Challenges}

\author{Mayira Sharif}
\affiliation{%
\institution{The University of Memphis}
\city{Memphis, TN}
\country{United States}}
\author{Guangzeng Han}
\affiliation{%
\institution{The University of Memphis}
\city{Memphis, TN}
\country{United States}}
\orcid{0009-0006-8355-2768}
\author{Weisi Liu}
\affiliation{%
\institution{The University of Memphis}
\city{Memphis, TN}
\country{United States}}
\orcid{0009-0004-8016-7587}
\author{Xiaolei Huang}
\affiliation{%
\institution{The University of Memphis}
\city{Memphis, TN}
\country{United States}}
\email{xiaolei.huang@memphis.edu}
\orcid{0000-0003-0478-8715}







\renewcommand{\shortauthors}{Sharif et al.}

\begin{abstract}

To support rapid AI advances and broaden access to large-scale computing resources for under-resourced institutions at the Mid-South, we established the first regional mid-scale GPU cluster at the University of Memphis (UofM), iTiger.
We present and analyze efforts of infrastructure management and computational support for educators, students, and researchers across scientific and engineering disciplines, such as precision agriculture, smart transportation, and health informatics.
We outline our initiatives to broaden cluster adoption on research and education, such as seed grant programs, workshop trainings, course integration, and other outreach activities.
We also identify challenges and further discuss findings of GPU infrastructure adoptions among college students and multidisciplinary researchers.
The insights will indicate how to effectively  and broaden infrastructure adoption and integrate into research and workforce developments.

\end{abstract}


  


\maketitle

\section{Introduction}


GPU cyberinfrastructure (CI) has been shown to have critical roles in AI research and education across disciplines, and broadening its adoptions among emerging research institutions remains a practical challenge ~\cite{Chakravorty_2024_BRICCs, rmacc}.
The Mid-South, including Arkansas, Mississippi, and West Tennessee and all centering on Memphis, has a mixture of R1 and R2 universities, primarily undergraduate institutions, and community colleges, many of which have limited access to the GPU resources.
In 2024, UofM established the first regional mid-scale GPU CI, iTiger\footnote{\url{https://itiger-cluster.github.io/}} supported by the National Science Foundation (NSF)\footnote{\url{https://www.nsf.gov/funding/opportunities/mri-major-research-instrumentation-program}}, which aims to promote CI adoptions in AI-related research and education across emerging research institutions.
While the GPU CI is still new to the regional institutions, a concrete question to be asked is: ``\textit{how can the regional GPU CI broaden access and stimulate research, education, and workforce development in the Mid-South?}''

Broadening GPU CI adoption for research and education is never an easy task among emerging research institutions. 
At premier research-intensive institutions, HPC usage is often deeply embedded in research culture, and students routinely access computing clusters through research projects and courses ~\cite{Brashear_2024}.
In contrast, the regional emerging research institutions provided only limited exposure to advanced GPU CI for researchers and students.
Many faculty and students have rarely used command-line interfaces or job schedulers on HPC systems, let alone optimized workloads for GPU acceleration. 
Compounding this, the regional practice has long lagged behind national trends in GPU CI adoption, with few institutions offering strong GPU resources prior to iTiger. 
These factors create a unique dual challenge: advancing research engagement while simultaneously building AI and HPC literacy across a largely computing-inexperienced student population. 
This raises a central question to be answered by this study: ``\textit{how can regional GPU CI lower barriers and promote CI literacy across teaching-oriented and under-resourced institutions?}''


To address the questions, we implemented several initiatives to support research development and foster undergraduate and graduate education through three major and complementary efforts: 1) research initiatives, 2) HPC training and workshops, and 3) curriculum integration.
Together, these efforts aim to cultivate a regional culture of CI engagement, community building, and workforce development, and thus promote CI adoption and literacy ~\cite{climatehpc, rmacc, hpc_pathways}.
Our research initiatives established seed-funding mechanisms to accelerate CI adoption and support researchers in developing new AI applications for scientific and engineering domains~\cite{jones2025examining, cai2025safetriage, imran2024vrpddtvehicleroutingproblem}.
Our training programs implemented regular tutorials and hands-on sessions to facilitate project development and deployment across a broad range of disciplines at the 10 regional institutions in Figure~\ref{fig:research}.
Finally, our education integration encouraged instructors to incorporate iTiger into course materials, including homework assignments and capstone projects in courses such as Introduction to Data Mining (COMP 4118/6118), enabling students to apply AI concepts in practice while strengthening their skills for research development.
This leads to a final question in this study: ``\textit{how effective are these strategies in promoting broad CI adoption, research engagement, and AI workforce development in the region?}''
We provide detailed analyses and discussions to address how CI adoption can be strengthened among emerging research institutions.

\begin{wrapfigure}{r}{0.58\textwidth}
    \centering
    \vspace{-15pt} 
    \includegraphics[width=\linewidth]{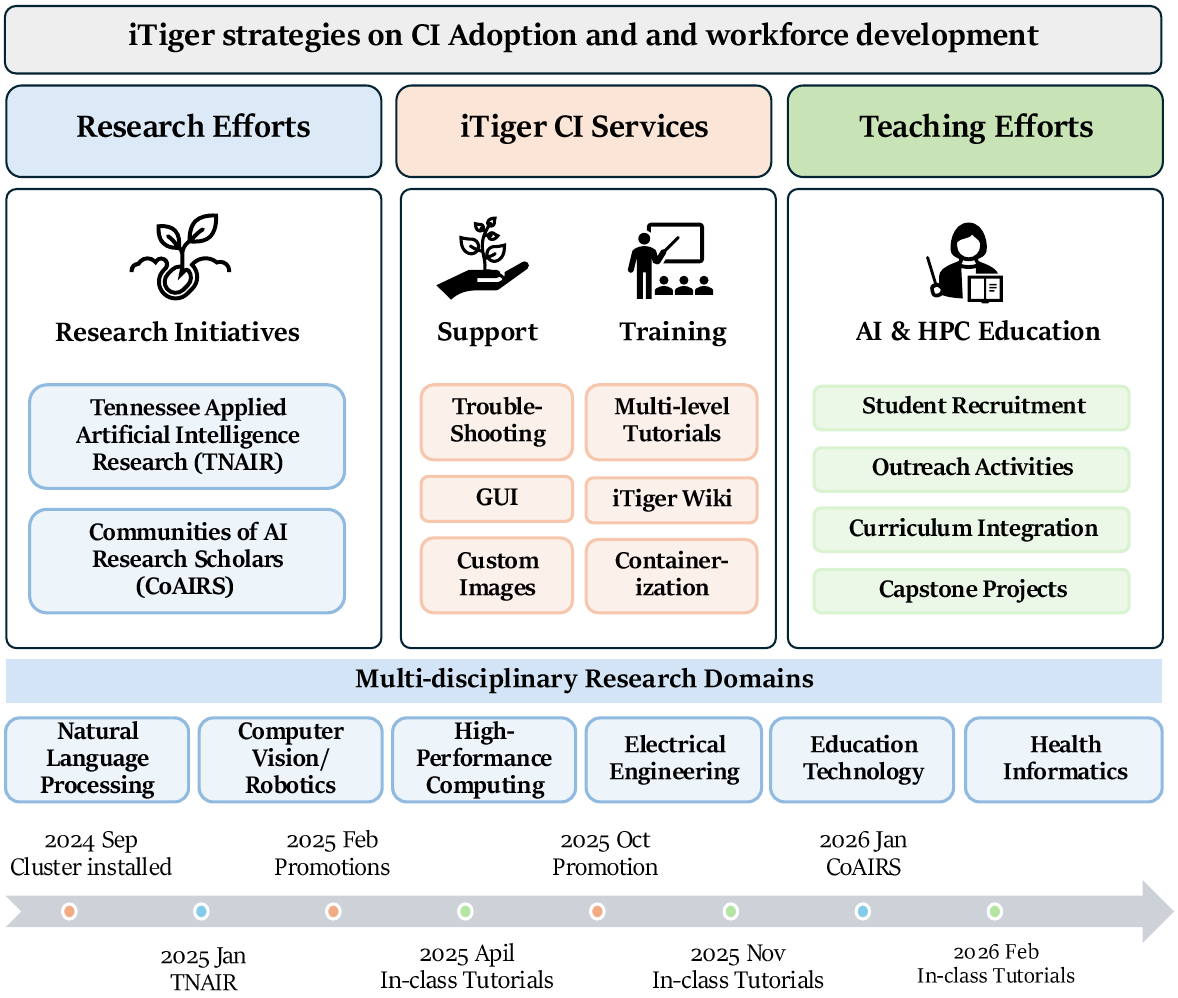}
    \vspace{-20pt} 
    \caption{Our efforts on CI adoption and workforce development.}
    \label{fig:strategy}
    \vspace{-20pt} 
\end{wrapfigure}
\section{Integrating iTiger CI for Regional Research and Education}



Since its setup in October 2024, the iTiger cluster has provided high-performance CPU/GPU computing resources for research and education in the Mid-South region. The cluster has experienced rapid user growth since its inception, now serving a community of 241 users with diverse backgrounds (Table~\ref{tab:user-demographics}). As illustrated in Figure~\ref{fig:strategy}, our work extends beyond iTiger CI services to actively drive regional research engagement and cultivate AI and HPC literacy among students and researchers. To achieve this, we complement our CI services with dedicated research efforts and teaching efforts, described in the following subsections.


\subsection{iTiger CI Services}

\paragraph{Tech Support}

On the support side, we have established multiple support channels to address issues ranging from basic access problems to complex trouble-shooting. Users can submit a contact request through our website for any questions, submit a ticket on our Wiki for any issues, or report an issue directly on our GitHub repository.
For users unfamiliar with command-line environments, we also deployed OpenOnDemand~\cite{Hudak2018,OpenOnDemand2023}, a web-based graphical user interface (GUI) that provides graphical access to the cluster without extensive technical training.
To support diverse computational environments, we provide custom container images pre-configured with AI acceleration libraries such as DeepSpeed and FlashAttention, so researchers can run advanced AI workloads without complex environment configuration. 

\begin{wraptable}{r}{0.6\textwidth}
\centering
\resizebox{.59\textwidth}{!}{
\begin{tabular}{ccc|cc|cc}
\toprule
\multicolumn{3}{c|}{Gender} & \multicolumn{2}{c|}{Status} & \multicolumn{2}{c}{Subject}  \\
\cmidrule(lr){1-3} \cmidrule(lr){4-5} \cmidrule(lr){6-7}
Unknown & Male & Female & Student & Faculty & STEM & Non-STEM  \\
\midrule
47 & 143 & 54 & 193 & 48 & 239 & 5   \\
\bottomrule
\end{tabular}%
}
\caption{\fontfamily{ptm}\selectfont iTiger CI user statistics.}
\label{tab:user-demographics}
\vspace{-1cm} 
\end{wraptable}

\paragraph{HPC Training}
To serve a broader range of users, we have developed platform-agnostic tutorials freely accessible to everyone regardless of their institutional status. 
These resources are categorized into three types: basic usage, model training, and parallel computing. 
The basic usage module covers essential skills such as remote connection protocols, fundamental command-line operations, Python virtual environment configuration, and container implementation. 
Currently, our tutorial repository includes seven comprehensive guides, supplemented by executable Jupyter Notebooks and beginner-friendly resources.
We are also actively developing more advanced tutorials focused on sophisticated deep learning model implementation and optimization techniques.




\subsection{Research Efforts}

\paragraph{Research Initiatives}
To accelerate CI adoption in research, we established seed funding initiatives at UofM. 
Our first initiative TN Applied Artificial Intelligence Research (TNAIR)\footnote{\url{https://www.memphis.edu/research/impact/newsletter_2024/december24_stories/tnair_call_for_proposals.php}} aims to promote adoption in multiple scientific disciplines, including transportation~\cite{imran2024vrpddtvehicleroutingproblem}, health, and engineering~\cite{wu2025design}. 
Then, we launched a new initiative Communities of AI Research Scholars (CoAIRS)\footnote{\url{https://memphis.infoready4.com/CompetitionSpace/\#competitionDetail/1995132}} to accelerate integrations across five strategic domains: Healthcare, Manufacturing, Software, Logistics, and Higher Education, reflecting growing interdisciplinary demands for CI resources.

\begin{wrapfigure}[10]{r}{0.63\textwidth}
    \vspace{-20pt}
    \centering
    \includegraphics[width=\linewidth]{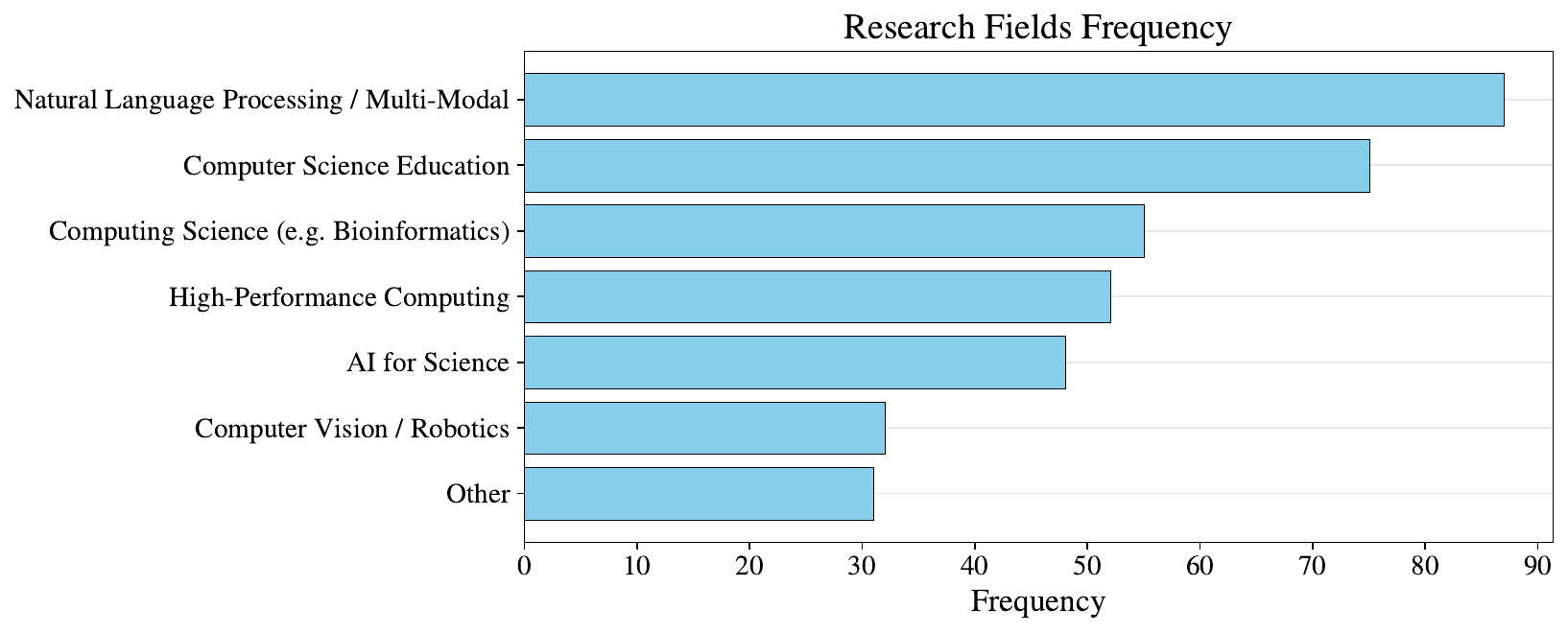}
    \vspace{-.7cm}
    \caption{\fontfamily{ptm}\selectfont Research fields frequency of iTiger from user surveys.}
    \label{fig:research}
\end{wrapfigure}

\paragraph{Multi-disciplinary Research}
iTiger supports a wide range of AI-related research fields, as evidenced by our user interest survey results (Figure~\ref{fig:research}). 
This cross-disciplinary usage pattern demonstrates the universal value of high-performance computing in modern research. 
The diverse applications of iTiger are reflected in actual user proposals spanning multiple domains. For example, iTiger CI supports AI-enhanced analysis of tumor progression prediction systems and next-generation AI-supported traffic control systems. 
Social science applications include analyzing trends in parental vaccine hesitancy on social media using natural language processing techniques, as well as projects examining mental health structures through computational methods. 
We have implemented specialized software stacks and toolchains, including domain-specific scientific computing libraries and analytical frameworks, to maximize cluster resource utilization and provide a valuable platform for cultivating interdisciplinary research talent. 
In addition, iTiger supports local conservation efforts, such as a collaboration with Memphis Zoo to develop models for endangered species protection.

\subsection{Teaching Efforts}

\paragraph{CI Student Recruitment}
To support both operational sustainability and workforce development, we have recruited two graduate students and one undergraduate student as cluster management assistants. 
Working alongside the University of Memphis's experienced research computing team, these students have accumulated 
practical experience in both HPC system usage and management.
These students play an active part in our website building, outreach activities, in-class tutorials, and workshop 
sessions. 
This approach not only creates valuable learning and professional development opportunities for the campus community, but also contributes to training CI professionals to strengthen long-term HPC workforce development in the Mid-South region.
\paragraph{Outreach Activities}
We have hosted a series of promotion activities to engage CI users through short 10-minute presentations delivered to students in various computer science courses, including both graduate-level and undergraduate-level courses. 
In these activities, we introduce how iTiger can support their coursework, research projects, and future career development,  and guide them through the account registration process to get started.
Thus far, we have presented to 151 students across 5 computer science courses.
\paragraph{Curriculum Integration and Capstone Project}
We actively integrate HPC resources into the curriculum and encourage students to use iTiger for both course projects and capstone projects.
For example, in COMP 4118/6118 (Introduction to Data Mining), we provide in-class sessions offering both HPC user guide and latest Machine Learning tutorials where students gain direct hands-on experience with the HPC cluster.
These practices create a mutually beneficial relationship between the curriculum and the cluster: 
students develop practical HPC skills and bridge the gap between machine learning concepts and real-world implementation, while their experiences inform ongoing improvements to the cluster's documentation and support systems.
\section{Results}
\begin{figure}[ht]
\centering
\vspace{-12pt}
\begin{minipage}{0.48\textwidth}
    \centering
    \includegraphics[width=\linewidth]{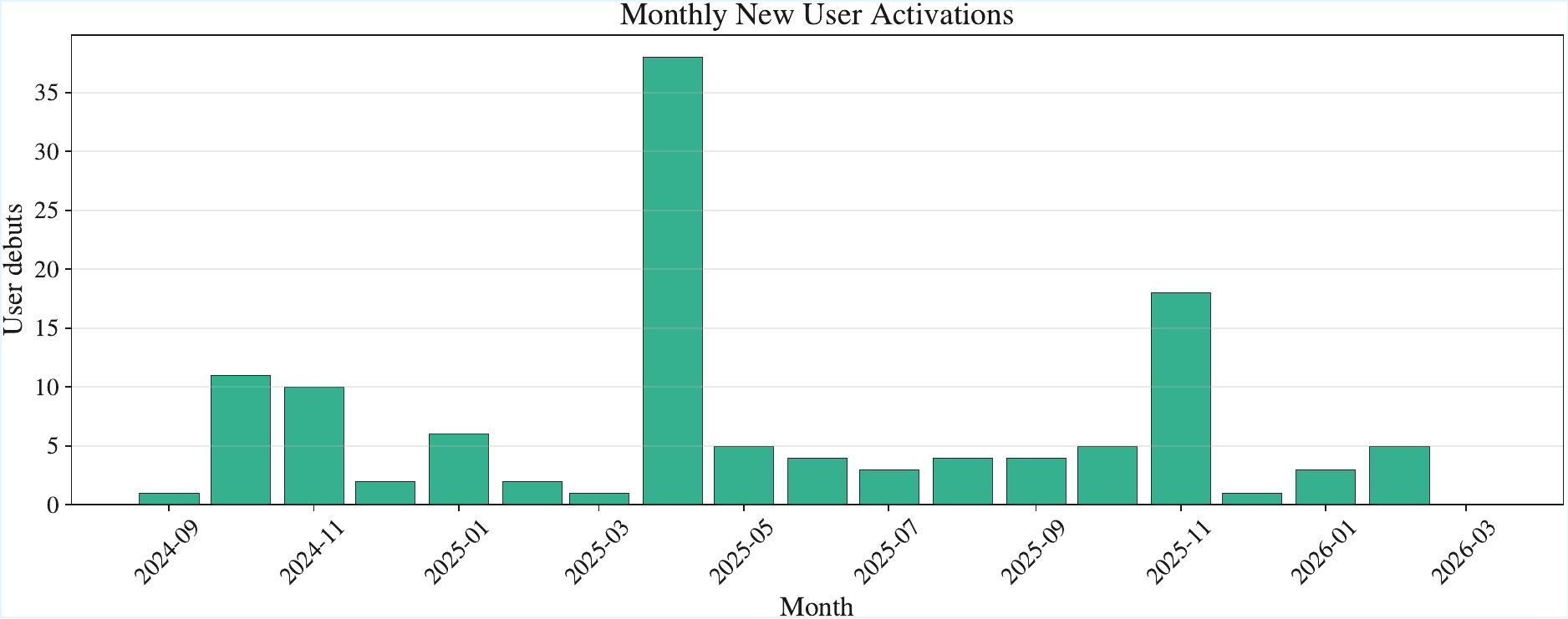}
    \caption{User debuts across time.}
    \label{fig:users}
\end{minipage}
\hfill
\begin{minipage}{0.45\textwidth}
    \centering
    \includegraphics[width=\linewidth]{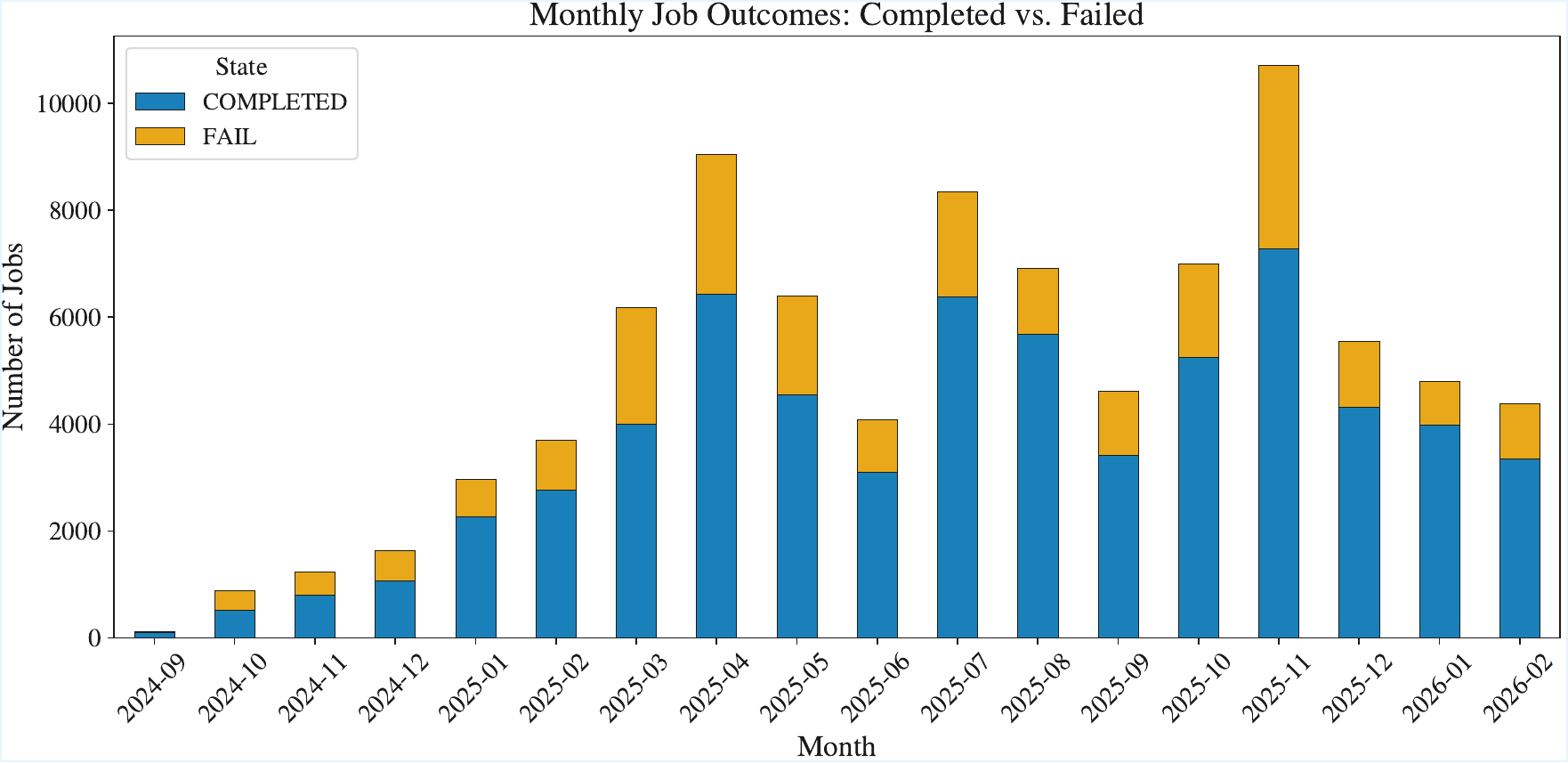}
    \vspace{-.7cm}
    \caption{Completed vs. failed job ratios over time.}
    \label{fig:complete}
\end{minipage}
\end{figure}

\paragraph{Training Outcomes}
To evaluate the short-term effects of our educational interventions, we collected cluster-level records on user onboarding and job execution behavior from September 2024 to February 2026. Figure~\ref{fig:users} shows that the new user debuts were not evenly distributed over time; instead, they concentrated on specific periods of instructional participation. 
In particular, April and November 2025, which correspond to our in-class tutorials and course-related training activities, exhibited clear increases in newly active users. 
These patterns suggest that structured, curriculum-connected exposure was an effective mechanism for introducing students to the cluster and lowering the initial barrier to CI adoption.

\vspace{-15pt} 
This growth in new users was accompanied by changes in job execution behavior. 
Figure~\ref{fig:complete} shows the number of failed jobs also increased during the same periods, while surrounding months exhibited substantially lower levels.
In this case, ``complete'' refer to jobs that finished without error, whereas ``failed'' means jobs assigned SLURM states such as FAIL, FAILED, and NODE\_FAIL. 
Rather than indicating a decline in system utility, these temporary increases in failed jobs are consistent with a phase of active experimentation, debugging, and workflow learning among newly onboard users.
The user majority were undergraduate students preparing end-of-course projects, even though their initial registration records alone did not fully reveal this trend.
Taken together, the temporal alignment between peaks in user debuts and shifts in job outcomes suggests that tutorial-based and course-integrated activities are critical in stimulating early hands-on engagement with iTiger.
These findings provide quantitative evidence that educational interventions can rapidly broaden CI adoptions, especially for users with limited prior HPC experience.

\vspace{-5pt}
\paragraph{Research Outcomes}

Beyond instructional usage, iTiger has supported a growing body of multidisciplinary scholarly outputs across regional institutions. These span health and clinical AI~\cite{jones2025examining, cai2025safetriage, li2025gin, xu2025enecg, xu2025ecgmoe}, agriculture and geoscience~\cite{hossain2025improving, bang2026application}, natural language processing~\cite{poria-huang-2025-bhaasha, han-etal-2025-attributes}, and smart transportation and engineering~\cite{imran2024vrpddtvehicleroutingproblem, wu2025design}. These outputs demonstrate that iTiger serves not only as an educational platform but also as a shared regional research infrastructure enabling publishable AI research across diverse scientific domains.

\vspace{-5pt}
\section{Discussion}

\paragraph{Integrating HPC into Curriculum is Critical for CI Adoption.}
Our experience shows that integrating HPC resources into coursework is an effective strategy for driving user growth and skill development. 
The peaks of new user debuts and job submissions occur in April and November 2025 in figure~\ref{fig:users} and figure~\ref{fig:complete}, confirming that course-embedded engagements are the primary driver of CI adoption among students. 
By encouraging instructors to incorporate 
iTiger into related courses (e.g. Data Mining), we enabled students
to apply AI concepts to real computational problems. 
The elevated job failure rates during these periods suggest active experimentation, which is a natural stage of HPC skill acquisition. 
These outcomes highlight that curriculum integration lowers the barrier to HPC entry and accelerates CI workforce development in the Mid-South region.

\paragraph{Research Initiatives Require Longer Time to Show Impact.}
Unlike our teaching efforts, the impact of research funding initiatives such as the Tennessee 
Applied Artificial Intelligence Research (TNAIR) program was not visibly reflected in aggregate user growth or job count trends. 
This is not entirely unexpected. 
Research projects typically 
involve extended preparation phases before any substantial computation begins, so their computational footprint may only materialize months after an award is made. 
Research adoption is also incremental by nature, making it difficult to isolate in short-term observations. 
Longer-term tracking will be necessary to more accurately assess the impact of seed funding on regional CI utilization.

\paragraph{Rethinking CS Education for HPC Integration.}
Through our efforts to integrate iTiger into capstone projects, we found that few or no existing capstone projects incorporated HPC or focused on AI applications.
Many current students entering these terminal-stage projects had limited exposure to HPC systems or AI applications in their prior coursework.
This gap is concerning given the growing demand for AI-competent graduates in the Mid-South region.
As the local economy increasingly shifts toward AI-driven industries, universities must ensure that graduates are equipped with relevant computational skills.
Current capstone projects, however, largely reflect outdated expectations that do not align with this evolving landscape.
We argue that HPC and AI literacy should be built progressively throughout the CS curriculum, so that capstone projects can serve as a culminating applied experience rather than a student's first encounter with these tools.
\section{Conclusion}
Our experience with iTiger demonstrates both the challenges and opportunities of establishing regional CI resources in historically under-served areas. While CI accessibility has improved across diverse Mid-South institutions, sustained development is needed in three directions: expanding cross-institutional research collaborations, integrating generative AI into user training for personalized guidance, and establishing pathway programs (e.g., internships and staff exchanges) with experienced HPC centers such as the BRICC program at Texas A\&M University~\cite{Chakravorty_2024_BRICCs}. These efforts can position iTiger not only as a computational resource but also as a catalyst for building sustainable CI capacity and addressing workforce shortages in the Mid-South region.

\bibliographystyle{ACM-Reference-Format}
\bibliography{main}
\end{document}